# Fast Homoepitaxy on (100) β-$Ga_2O_3$ Substrates with Large Grown-In Offcut

M. Brooks Tellekamp,[1*] Drew Haven,[2] David Joyce,[2] John Mangum,[1] Henry Garland,[1] Robert Lavelle,[3] Kevin Schulte,[1] Anna Sacchi,[1] Matt Young,[1] Andriy Zakutayev[1]

[1]National Laboratory of the Rockies, Golden CO, USA

[2]Luxium Solutions, Millford NH, USA

[3] Pennsylvania State University Applied Research Laboratory, State College PA, USA

The choice of crystalline orientation and offcut angle is non-trivial for low symmetry monoclinic β-$Ga_2O_3$, where anisotropy impacts bulk and thin film synthesis, material properties, and power device fabrication and performance. Scalable (100)-oriented β-$Ga_2O_3$ wafers are desirable for electronic devices because the theoretical breakdown field is larger than other orientations, but it is not typically used for devices due to 10-30x slower epitaxial growth rates compared to other orientations. Here we report growth rates up to 5.1 nm/min, equal to rates obtained on the fastest growth rate (010) face, by using (100) $Ga_2O_3$ wafers with large grown-in offcuts. The offcuts (3.4° - 13.4°) are directly grown by Edge-defined Film-fed Growth (EFG) of 2D ribbons with rotated seed crystals, avoiding material loss in typical 3D crystal boule offcut methods while maintaining high crystalline quality. Chemical-mechanical polishing and dicing of the ribbons into substrates resulted in surfaces with roughness ≤ 0.2 nm RMS. The surface morphology of MBE-grown layers at 650 °C substrate temperature shows (100) terraces and ($\bar{2}01$) step edges, evidence of step-flow growth, are maintained at these high offcuts with step-bunching above 700 °C. Electrochemical capacitance-voltage profiling gives an unintentional n-type doping density of $2\times10^{15}$ $cm^{-3}$ – $7\times10^{15}$ $cm^{-3}$, one of the lowest values reported for MBE-grown films, due to the increased growth rate. Planar Schottky barrier diodes on unintentionally doped MBE epilayers without edge termination have an on/off ratio ~$10^5$ and an average breakdown field of 1.56 MV/cm on a 13.4° offcut (100) surface, comparable to devices with similar architecture fabricated on other crystal faces. Overall, these results illustrate co-design of the crystal face and offcut angle for low-symmetry $Ga_2O_3$ crystals and validate the use of the scalable (100)-oriented β-$Ga_2O_3$ wafer for fast epitaxial growth and electronic device fabrication.

## Introduction

The choice of crystalline orientation impacts crystal growth, epitaxial growth, properties, and device design, and this impact increases in complexity as crystal symmetry reduces. For example, the choice of As- or Ga-terminated cubic GaAs surfaces, (111)-A and (111)-B respectively, results in amphoteric Si doping and anisotropic bulk growth.[1] Lower symmetry hexagonal systems like GaN introduce anisotropic polarity which can be leveraged in the case of a 2-dimensional electron gas,[2] or which must be mitigated in the case of light-emitting diodes (LEDs).[3] Additionally, offcut crystals can modify epitaxy by breaking surface symmetry and promoting step-flow growth. Offcut angle and direction is critical in SiC epitaxy, where the offcut can effectively control 4H, 6H, and 3C polytype formation.[4] Monoclinic β-$Ga_2O_3$ has greater crystal anisotropy, leading to more choices in both substrate orientation and offcut angle.

β-$Ga_2O_3$ is an important ultra-wide bandgap (UWBG) semiconductor for power electronics and extreme environment electronics applications with rapidly growing interest. The first β-$Ga_2O_3$ transistor was demonstrated in 2012,[5] yet ampere-class diodes were commercialized in 2022 and 6" substrates are now commercially available.[6] The rapid growth of $Ga_2O_3$ (all uses of $Ga_2O_3$ in this manuscript refer to the monoclinic β-phase) is driven by both promised performance and economics. $Ga_2O_3$ is predicted as one of the highest performing power semiconductor due to an ultra-wide bandgap (4.6 eV), sufficient n-type carrier mobility (~200 $cm^2$/V-s at 300 K),[7–9] and shallow doping (36 meV isolated donor activation energy),[10] leading to high figures of merit for power handling and switching.[9] Due to the ease and scalability of $Ga_2O_3$ crystal growth, it is forecast to cost 5x-10x less than SiC based on conservative estimates for iridium crucible reuse,[11,12] and efforts to reduce iridium consumption in crucibles will lead to further cost savings.[13]

The pace of $Ga_2O_3$ scaling and device demonstrations has been rapid, and the field quickly converged to (001)-oriented crystals due to bulk scalability and ease of drift layer epitaxy through a technique called halide vapor phase epitaxy (HVPE). However, recent research suggests that alternative crystal orientations may be more favorable for wholistic co-design of $Ga_2O_3$ growth, properties, and device fabrication. These findings have renewed interest in alternative substrate orientations such as (100), both for improved scalability and greater performance. While originally abandoned due to slow growth and high density twin-domains, it is now well-known that (100)-oriented wafers offcut in the [-c] direction suppress twin domain formation, and that the offcut promotes step-flow growth and increases the growth rate with maximum offcut angle of 6°.[14–17] Greater offcuts have not yet been studied due to the waste associated with cutting, grinding and polishing large offcuts into boule-grown crystals.

Here, we report on $Ga_2O_3$ homoepitaxy on highly offcut (100)-oriented substrates, with grown-in offcuts up to 13.4° in the -c direction, produced by Edge-defined Film-fed Growth (EFG).[18] The impact of grown-in offcut angle on growth rate and surface morphology is explored by MBE, and the growth-rate continues to increase up to the maximum-explored offcut angle of 13.4 °. We find our comparative "ideal" growth rate on (010)-oriented substrates, 4.0 – 4.5 nm/min (Figure S1), is matched or exceeded on highly offcut (100) substrates and this growth rate is equal to previous MOCVD reports (4.3 nm/min) on 6° offcut (100) substrates,[19] noting that optimized MOCVD growth on offcut (100) $Ga_2O_3$ has exceeded 16 nm/min.[16] Electrochemical capacitance voltage profiling shows the unintentional doping (UID) density of our MBE-grown films < $2×10^{15}$ $cm^2$, the lowest UID concentration reported for MBE-grown $Ga_2O_3$ to date. Vertical Schottky barrier diodes (SBDs) fabricated on 550 nm MBE-grown epilayers without field management or edge termination demonstrate an average breakdown field of 1.56 MV/cm, comparable with similarly fabricated SBDs on (001) $Ga_2O_3$.

## Co-Design of Anisotropic Properties

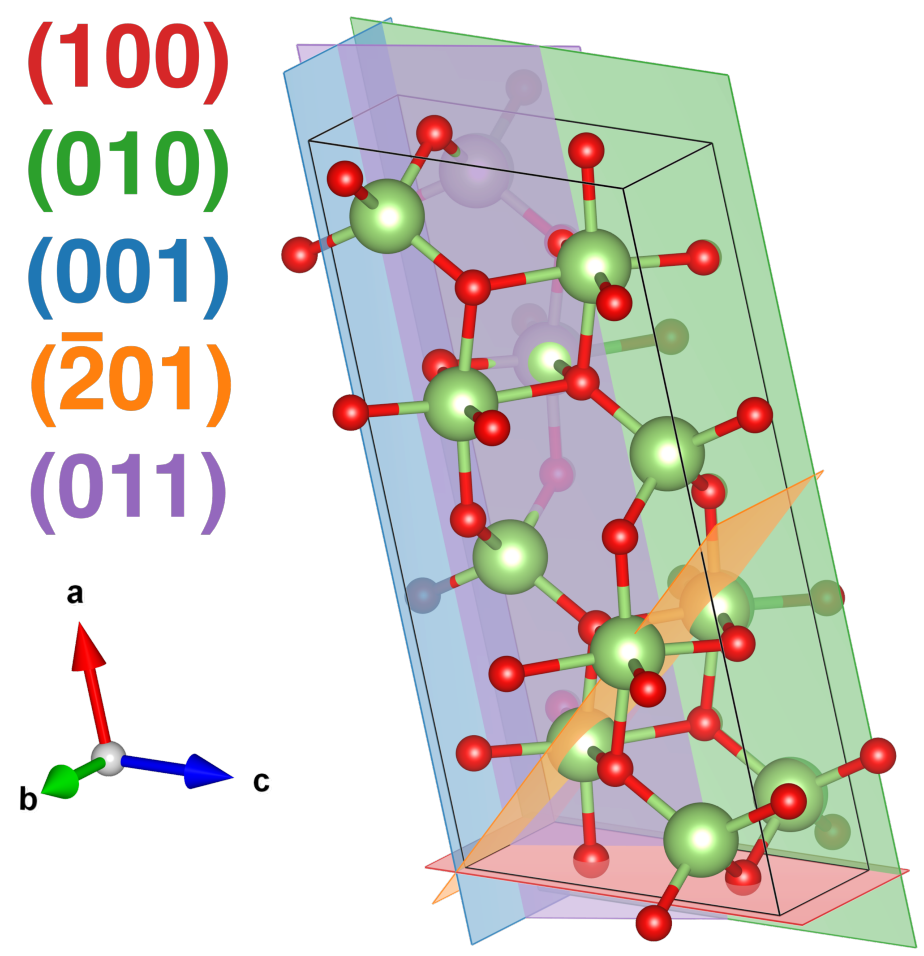

*Figure 1. Commercially available orientations of monoclinic β-$Ga_2O_3$. Crystal orientations have outsized impacts on bulk growth, substrate preparation, epitaxial growth, material properties, and device design for this low-symmetry crystal.*

First, we consider the primary commercially-available orientations of $Ga_2O_3$ in context of currently known properties (Table 1). Figure 1 shows the most common commercially-available orientations of $Ga_2O_3$, noting that (110) and

*Table 1. **Anisotropic properties of $Ga_2O_3$.** Considerations at the material properties, bulk substrate growth, and epitaxy level for commercially available faces of $Ga_2O_3$, corresponding to the crystalline faces shown in Figure 1.*

| | | (001) | (100) | (010) | (-201) | (011) |
|---|---|---|---|---|---|---|
| Properties | Theoretical Breakdown Field (MV/cm)[20] | 7.6 | 10.2 | 4.8 | n.d. | n.d. |
| | Electron Effective Mass ($m_e^*$)[21,22] | Nearly isotropic ~0.28 | | | | |
| | DC Dielectric Constant ($\epsilon_r$)[23,24] | 12.4 | 10.2 | 10.6 | est. 11.4 | est. 12 |
| | Optical Band Gap (eV)[25–27] | 4.7 | 4.7 | 4.54-4.57 | 4.52 | n.d. |
| | Thermal Conductivity (W/m-K)[28–30] | 12.7 - 13.7 | 9.5 – 10.9 | 22.5 - 23.4 | ~11.4 | ~20.3 |
| Substrates | Defect Types[15,31,32,32–39] | twin, groove, line | twin | Screw, pipe, line, pit | Edge, line, twin | Poly |
| | Large Area Scalability[31,40–44] | Yes | Yes | No | Yes | Yes |
| | Cleavage Plane[15,38,44,45] | yes | yes | No | No | no |
| | Substrate Growth maturity[40,41,43,44] | Yes | No | Yes | No | No |
| Epitaxy | Epitaxial Surface Roughness[15,45,46] | Good | Good | Good | Medium | Good |
| | Epitaxial Growth Rate[44,45,47–50] | Medium | Slow | Fast | Medium | n.d. |
| | Coincident lattice with p-type oxide[51–55] | (331) cubic | (100) cubic | (110) cubic | hcp | n.d. |
| | HVPE maturity[43,44,49,50] | Yes | No | No | No | No |

*n.d. = no data, est = estimated, hcp = hexagonal close-packed*

(101) have also been reported, and (101) is a close-packed surface pseudo-equivalent to ($\bar{2}$01).[27,47,56] Very recently, *Oshima et al.* demonstrated that the oxygen sublattice of $Ga_2O$ can be represented as a pseudocube, further motivating the analysis of alternative faces that are pseudo-equivalent to cubic (100), (110), and (111) orientations.[57]

As (001)-oriented technology has advanced, research has shown that impurity concentrations are higher than other orientations and has also noted the presence of line defects, twin domain formation, and pits.[31,32] The (010) and (-201) faces also demonstrate multiple types of extended dislocations.[58] Notably – the (100) orientation is considered to be 'dislocation-free', with the exception of twins, because all dislocations are parallel to this cleavage plane.[59] Additionally, due to anisotropy in electron-electron interactions, the [100] direction in $Ga_2O_3$ is predicted to have a significantly lower impact ionization coefficient and therefore a greater critical field for avalanche breakdown – 1.3x greater than [001] and 2x greater than [010].[20] Experimental support for devices utilizing the [100] direction has recently been published; the authors report a 1.8x – 2.7x increase in breakdown field when comparing vertical heterojunctions on (100)-oriented and (001)-oriented substrates, with a highest reported breakdown field for the [100] direction of 7.5 MV/cm.[51] NiO and $Cr_2O_3$ p-n heterojunction performance on $Ga_2O_3$ has also been reported as highly anisotropic, likely due to the heteroepitaxial relationship between the two materials.[51,60] Finally, while too cumbersome to include in the table, device fabrication concerns are equally important for substrate orientation choice. For example, the common recessed fin architecture shows anisotropic performance when the fin sidewalls are (100) or (010) surfaces, with (100) sidewalls showing lower charge-trapping than (010) sidewalls.[61] It has also been shown that plasma etching damage and wet etching are both highly anisotropic.[62,63]

Reported $Ga_2O_3$ growth rates for conventional plasma-assisted MBE are 2 – 2.5 nm/min for (010) oriented crystals,[64,65] 0.7 – 1.4 nm/min for (001)-oriented crystals,[65–68] and 0.1 – 0.15 nm/min for (100)-oriented crystals.[67] There has been significant research effort to increase these growth rates through multiple methods. Ozone-assisted MBE has reported up to 10.8 nm/min on (010)-oriented crystals.[69] Suboxide MBE (using $Ga_2O$ as a solid source) has reported rates up to 25 nm/min in ozone on (010)-oriented crystals.[70] The most studied method is indium-catalyzed MBE which has reported rates up to 5 nm/min on (010)- and ($\bar{2}$01)-oriented crystals,[65] and 0.33 nm/min on (100)-oriented crystals.[34] The results here represent a significant step forward in $Ga_2O_3$ growth, the highest offcuts provide a 3.4x increase in growth rate compared to the prior best reports on (100) offcut substrates.

## Results and Discussion

## *Rotated Seed Crystal Growth*

To enable exploration of extremely offcut surfaces without significant grinding and polishing losses, β-$Ga_2O_3$ crystals were grown by the EFG technique with a rotated seed.[18,71] Orientations for the pulled ribbons were (100) with a rotation toward [00$\bar{1}$], pulled in the [010] direction. Direct rotation of the seed crystal enables access to extreme offcut angles without traditional grinding and polishing losses. Rotations toward [00$\bar{1}$] ranged from 3.4° to 13.4°. This produces a ribbon with a grown in miscut that does not have to be machined in post growth. The direct-offcut EFG method is shown in Figure 2a, and an example Sn-doped ribbon with a 6° grown-in offcut before polishing is shown in Figure 2b. After bulk synthesis, an optimized chemical-mechanical polishing (CMP) routine was used to produce an epi-ready surface. A polished and diced wafer with an 11.1° grown-in offcut is shown in Figure 2c.

After final CMP, the offcut angle and overall crystalline quality of the ribbons were assessed using HRXRD. (400) full-width at half-maximum (FWHM) values < 100" were measured when orienting the beam parallel to <010> planes. The surface topography was also characterized using AFM, and an example 500x500 nm scan of a 12.8° offcut wafer is included in Figure 2d, showing a surface roughness of 0.15 nm RMS. An additional scan of a 5.7° offcut wafer is shown in Figure S2. The average terrace width decreased with increasing offcut angle. The samples were wet etched and annealed before growth to expose terraces, and the surface roughness was on the order of 0.3 nm RMS after wet etching and 0.6 nm RMS after annealing.

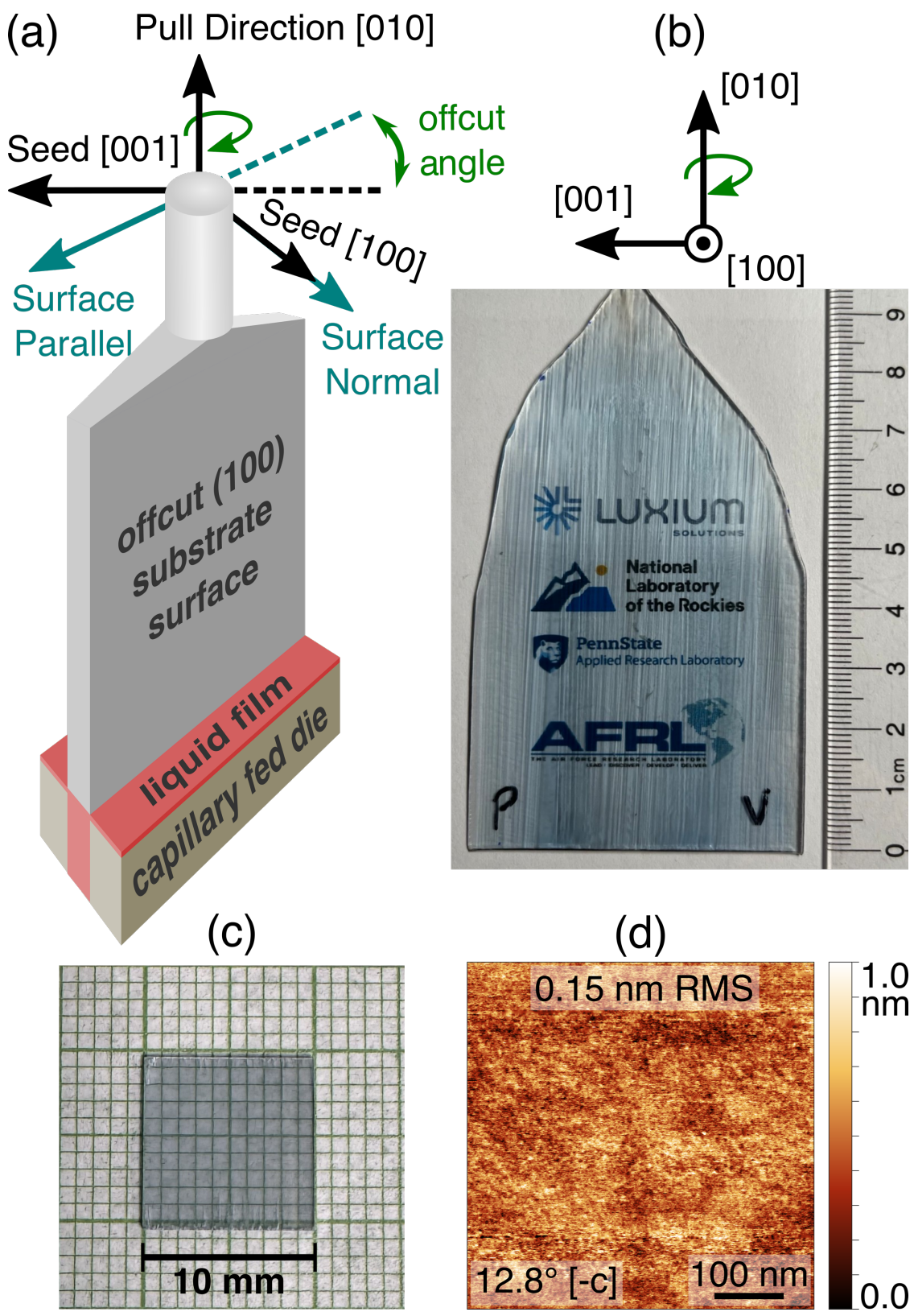

*Figure 2. **Grown-in offcuts of varying angle with Edge-defined Film-fed Growth (EFG).** (a) The grown-in offcut is obtained by pulling the ribbon through the die with a rotated seed crystal, shown here with a 13° rotation ~equal to the β angle. (b) (100) Sn:$Ga_2O_3$ ribbon with a 6° grown-in offcut, 55 mm wide. (c) (100) Sn:$Ga_2O_3$ ribbon with an 11.1° grown-in offcut, polished and diced to 10x10 mm. (d) AFM surface image of a 12.8° offcut (100) $Ga_2O_3$ wafer after polishing and cutting.*

## *Epitaxial Growth*

To explore the impact of grown-in offcut angle on growth rate, $Ga_2O_3$ epilayers were grown for 30 minutes on top of a thin $(Al,Ga)_2O_3$ phasor while varying substrate temperature and substrate offcut. The $(Al,Ga)_2O_3$ phasor serves as a density discontinuity to promote observation of Laue oscillations by XRD. For this experiment the Ga flux was fixed at $1\times10^{-6}$ torr BEP, which was calibrated for [010] oriented $Ga_2O_3$ to be slightly oxygen-rich. Measured growth rates are shown in Figure 3a for substrate temperatures ranging from 600 °C to 750 °C, and an [00$\bar{1}$] offcut angle ranging from 3.4° to 13.4°. Exemplary XRD is shown in Figure 3b for a series of epilayers grown for 30 minutes on increasingly offcut substrates, where the substrate temperature was fixed at 700 °C. For all substrate temperatures investigated, the growth rate is observed to increase with greater offcut angle. For comparison, the measured growth rate for [010] oriented $Ga_2O_3$ at 650 °C with the same Ga flux – $1\times10^{-6}$ torr BEP – was 4.1 nm/min (Figure S1). This growth rate is nearly identical to the film grown at 650 °C on the 13.4° offcut substrate – 4.16 nm/min. Of note, the 13.4° offcut is very close to the deviation of the β-angle (103.8°) from normal – 13.8°, bringing the a-oriented atomic columns normal to the surface.

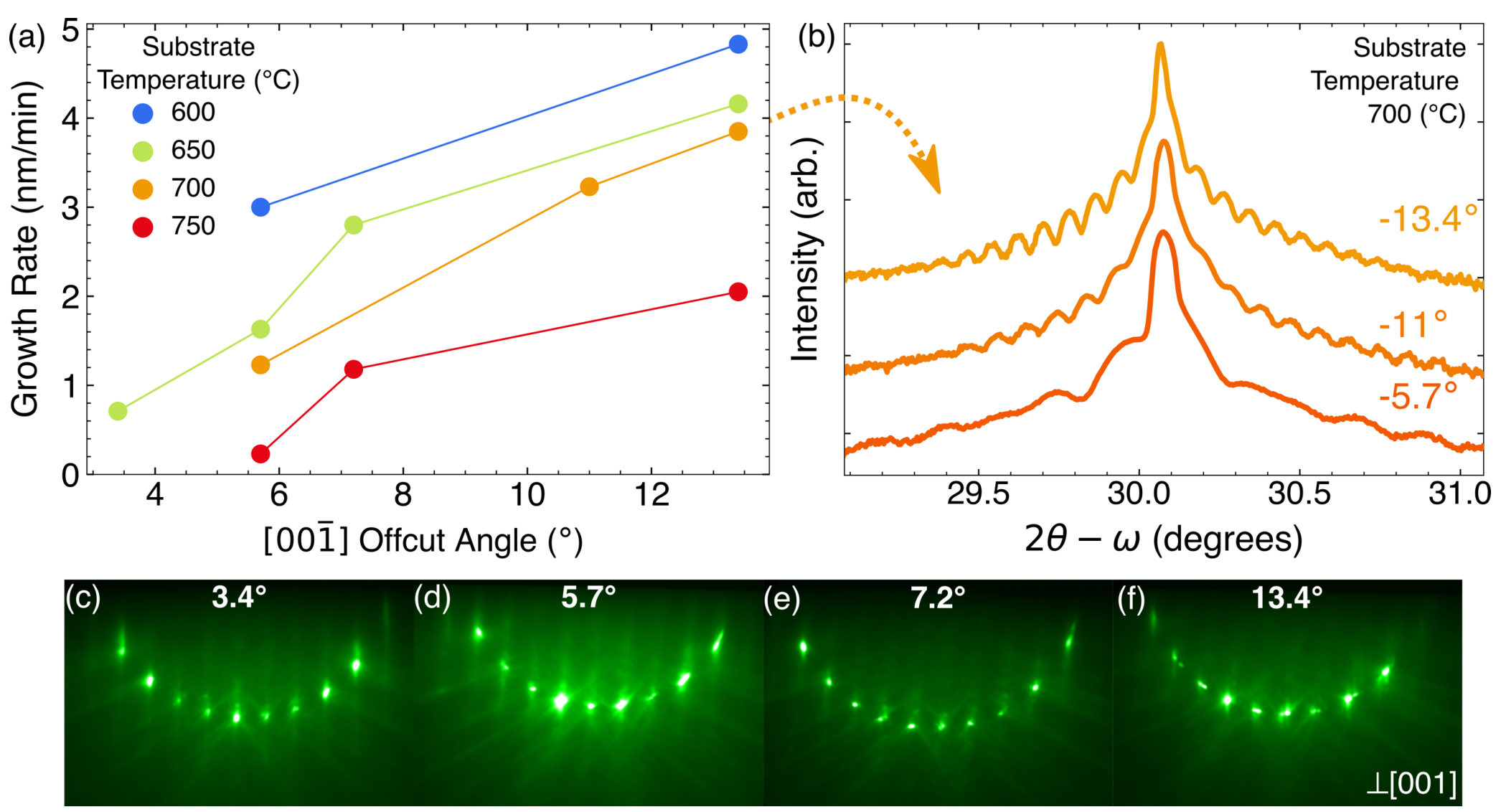

*Figure 3. **Growth rates of $Ga_2O_3$ increase with offcut angle and decrease with substrate temperature. (a)** Growth rate of $Ga_2O_3$ as a function of substrate temperature and offcut angle towards the -c direction, measured by XRD thickness fringes. The growth rate continues to increase up to a maximum measured offcut of -13.4°. **(b)** XRD 2θ-ω symmetric scans of $Ga_2O_3$ epilayers grown for 30 minutes at 700 °C show increasing thickness for greater offcut angles. **(c-f)** As-loaded RHEED patterns of $Ga_2O_3$ substrates offcut by 3.4°, 5.7°, 7.2°, and 13.4° show [001] characteristic diffraction with streaks inclined by the offcut angle. For higher offcuts (d-f) multiple reflections are observed within each diffraction spot corresponding to diffraction from parallel terraces, most clearly observed in (f).*

RHEED was used to evaluate the surface quality of substrates before growth, and to monitor the progression of epilayer growth. Figure 3c-f shows representative RHEED images of as-loaded substrates with offcut angles 3.4°, 5.7°, 7.2°, and 13.4°. The diffraction patterns of each substrate are representative of the same diffracting surface – the electron beam is perpendicular to [001] producing streaks that are separated by 1/c in reciprocal lattice units. Due to the offcut angle, the diffraction streaks are tilted, and the observed diffraction streak tilt angle increases as the offcut increases. For the higher offcut angles, fine structure is visible within each diffraction spot. Splitting is observed within the primary diffraction streaks, and these streaks are oriented vertically. This pattern indicates diffraction from an ordered step-terraced surface, where the diffraction from evenly spaced terraces creates the fine structure. This observation indicates that even for extreme offcuts of 13.4° the growth surface is still terminated by (100) planes. If we assume that each terrace step is one-half unit cell in height (6.1 Å), the expected terrace width for the highest offcut substrate, 13.4°, is 2.3 nm – 2.9 nm in the [001] direction. This highly stepped surface requires a reduction in adatom diffusion length to prevent step-bunching. The most straightforward method to reduce adatom diffusion length is to reduce the growth temperature, which also increases growth rate by reducing $Ga_2O$ desorption. Not only do increased substrate offcuts improve the growth rate; they also promote growth at conditions which further increase the growth rate.

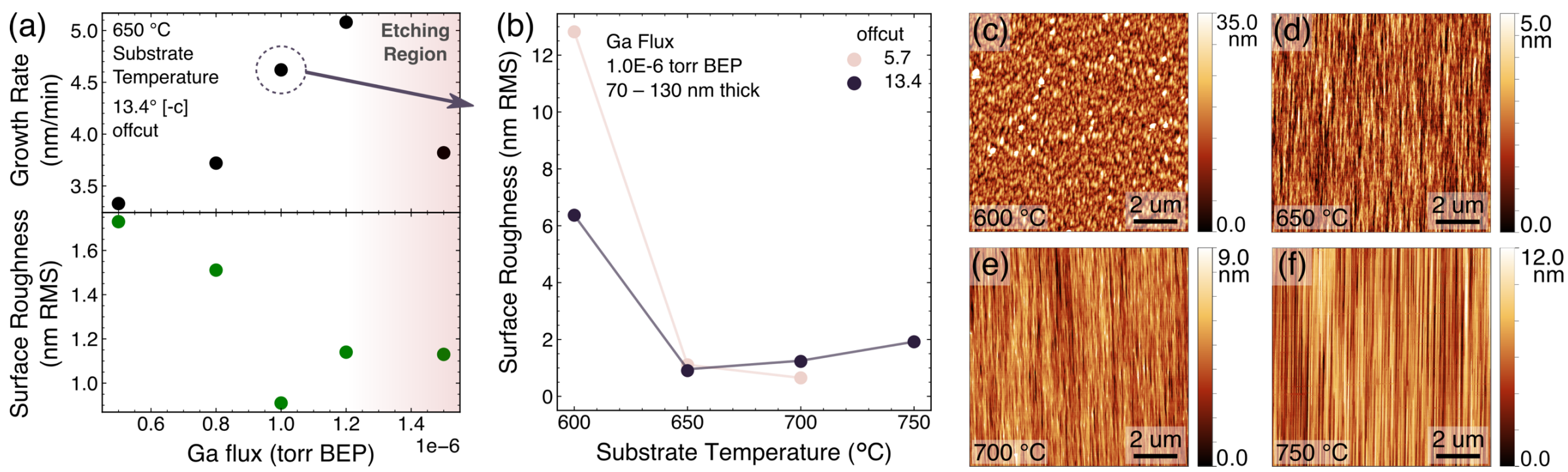

*Figure 4.* ***Impact of offcut and growth condition on surface morphology. (a)*** *Growth rate and surface roughness as a function of Ga flux for a fixed condition of 650 °C substrate temperature and 13.4° [-c] offcut substrates. The onset of etching is observed, corresponding to a transition from O-rich to Ga-rich growth.* ***(b)*** *The surface roughness extracted by AFM for 5.7° and 13.4° [-c] offcuts versus growth temperatures, limited to films 50 nm – 150 nm in thickness. For the 5.7° offcut the smoothest surface is obtained at 700 °C, while for the 13.4° offcut 650 °C results in a smoother film due to the reduced terrace length associated with larger offcuts.* ***(c-f)*** *AFM scans for the 13.4° offcut datapoints in (b). Step-bunching is observed for the film grown at 750 °C. For both 5.7° and 13.4° offcuts the films grown at 600 °C show a grainy structure indicating insufficient adatom diffusion.*

## Surface Morphology and Growth Mode

The impact of growth condition on surface morphology was investigated by tapping-mode AFM. First, a sample set was grown (targeting 100 nm thick) varying by the Ga flux, and therefore III:VI ratio, while at constant substrate temperature and offcut angle, 650 °C and 13.4° [-c] respectively. The growth rate and surface roughness of these films are shown in Figure 4a, and the AFM images are shown in Figure S3. The onset of gallium suboxide desorption, evidenced by a decreased growth rate, is seen above $1.2\times10^{-6}$ torr BEP Ga, and the maximum obtained growth rate is 5.08 nm/min. This maximum growth rate is faster than other reports for conventional plasma-assisted MBE on any orientation, and is also equal to or exceeding reported rates for indium-catalyzed $Ga_2O_3$ on any orientation.[65] The surface roughness was minimized for slightly oxygen-rich films, and growth in the etching regime did not impact the surface morphology. Growths performed with 8×10$^{-7}$ torr BEP Ga or less demonstrated triangular surface features consistent with (012) facets. Previous reports of chemical vapor deposited $Ga_2O_3$ on offcut (100)-oriented wafers have reported step pinning in oxygen-rich conditions related to higher concentrations of silicon which is mitigated by increasing the relative Ga precursor flow rate.[72] Due to the unavoidable presence of silicon at the interface and use of silica powder for chemical-mechanical polishing, the mechanism we observe is likely similar.[73] To minimize surface roughness, the 1×10$^{-6}$ torr BEP condition was chosen for further study.

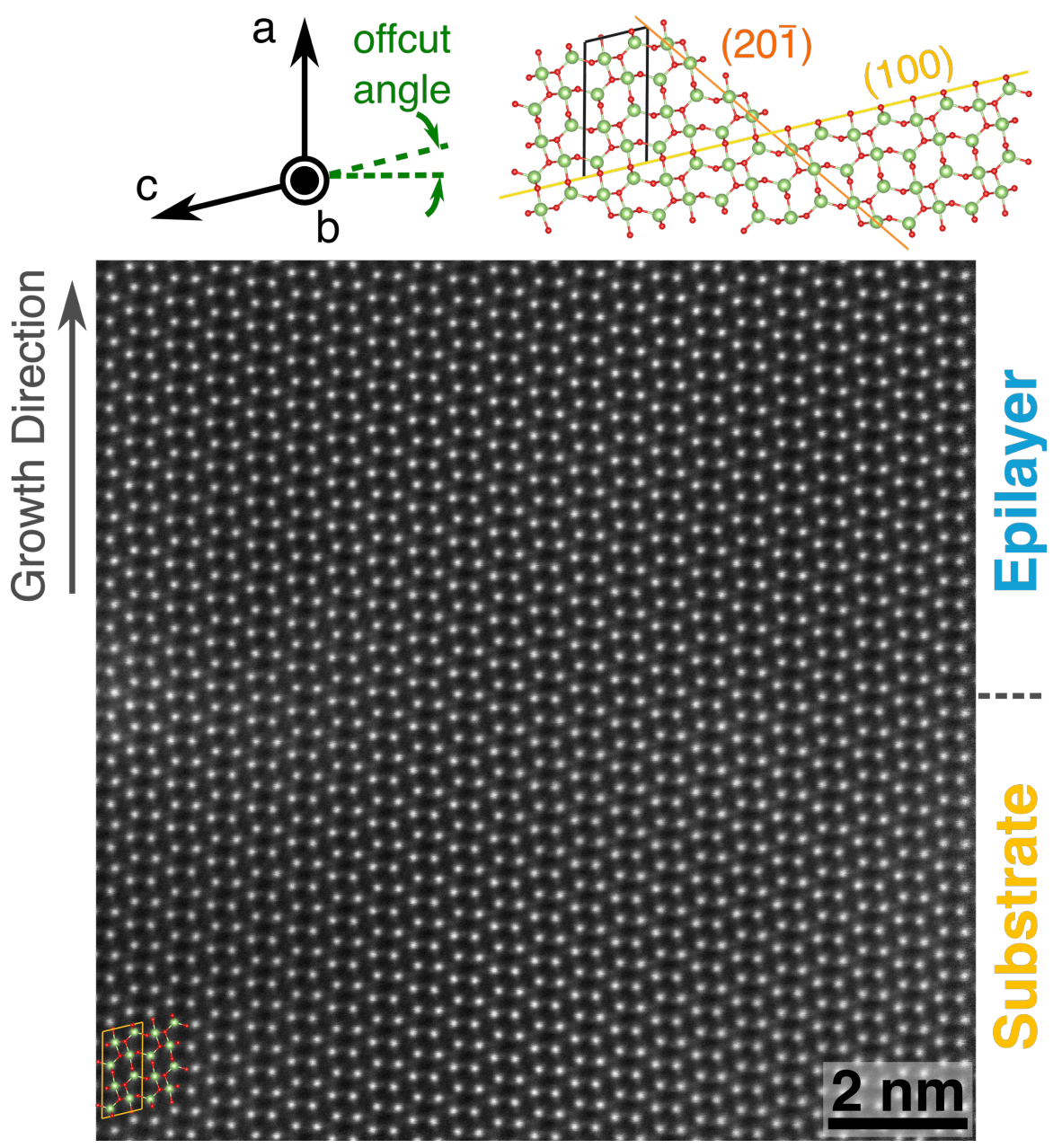

*Figure 5.* ***Microscopy of a highly offcut interface. Cross-sectional*** *STEM HAADF image acquired at the substrate-epilayer interface of a 13.4° offcut sample. The image is vertically oriented with the growth direction, and the offcut is visualized on the coordinate diagram. The crystal structure model shows an expected (100) terrace with a ($\bar{2}$01) facet marked in orange.*

Films were grown on 5.7° and 13.4° offcut substrates at varying substrate temperature to explore the transition from step-flow to step-bunched growth. For the 13.4° offcut films, the targeted thickness was 100 nm. For the 5.7° offcut films the thickness ranges from 50-100 nm. The surface roughness is presented as a function of substrate temperature and offcut angle in Figure 4b. For the 5.7° offcut, the surface roughness decreases with increasing temperature up to 700 °C (the 750 °C film is too thin for meaningful analysis due to the low growth rate). AFM images from the 5.7° offcut series are shown in Figure S4. The 13.4° offcut samples show gradually increasing roughness above 650 °C. At the higher offcut, the smaller terrace width promotes step bunching at lower temperatures than observed at lower offcuts, resulting in increased roughness. A comparative sample from the 11° offcut batch showed similar surface morphology to the 13.4° offcut wafer. Figure 4c-f show the AFM data for the 13.4° offcut series in Figure 4b. At 600 °C, there is insufficient surface adatom diffusion to promote a smooth film, and islands form. The same morphology was observed for 5.7° offcut films grown at 600 °C (Figure S4). Above 600°C, striations along [010] are observed corresponding to the offcut-determined step terraces. Films grown at 650 °C and 700 °C are qualitatively similar with slightly increased roughness, with clear signatures of step-bunching and increased roughness at a substrate temperature of 750 °C.

(100) substrates with 6° offcuts in the $[00\bar{1}]$ direction are known to expose $(\bar{2}01)$ facets which seed step-flow growth, while 6° offcuts in the [001] direction expose (001)-B terraces on which it is favorable to nucleate an unfavorable (101) twin-boundary.[15] It is unknown, at these extreme offcuts, if the same $(\bar{2}01)$ facet will be exposed, and if twin domains will form at the interface. Scanning transmission electron microscopy (STEM) was performed on the 13.4° offcut sample grown at 650 °C to visualize any interface defects associated with the high density of step terraces. A high-resolution high angle annular dark field (HAADF) STEM image of the substrate-epilayer interface is shown in Figure 5. The image is vertically aligned with the growth direction, therefore the [100] atomic columns are nearly vertical due to the offcut angle of 13.4° which is nearly identical to the β-angle inclination from 90°. The interface was located from a wider area scan where some substrate surface impurities were identified related to the cleaning procedure (Figure S5). The images indicate that the substrate-epilayer interface is pristine. An 800-nm wide HAADF image (Figure S5) confirms the layer is uniform, and the epilayer surface has maintained a step-and-terrace structure with (100) terraces and $(\bar{2}01)$ step-edges. AFM of a 600 nm thick epilayer grown on an 11.1° offcut substrate still maintains step-bunched terraces with a 1.34 nm RMS surface roughness (Figure S6), indicating the growth mode is maintained over significant thickness. These results corroborate a step-flow growth mode, although step-bunching was not eliminated and further optimization to the growth conditions could lead to even further improvement.

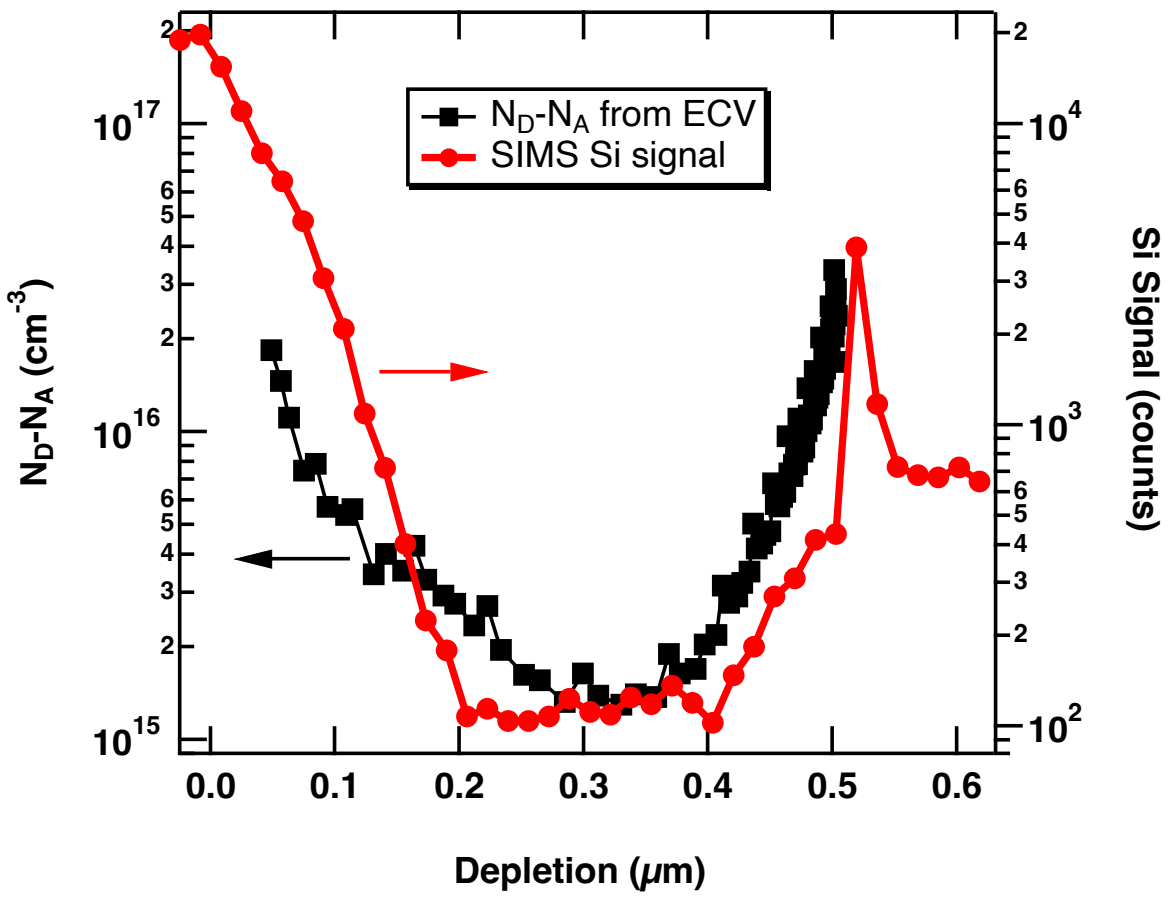

*Figure 6.* ***Electrochemical capacitance-voltage (ECV) profiling of thick MBE-grown epilayers.*** *The unintentional doping concentration $N_D$-$N_A$, as measured by ECV, is less than 2E15 $cm^{-3}$. SIMS Si counts are displayed on the same x-axis, the flat region representing the detection limit.*

## *Electrical characterization*

Electronic devices require strict control over unintentional doping densities to achieve the greatest electric field without breakdown. To measure the unintentional doping density of the MBE-grown epilayers on 13.4° offcut substrates, we grew a 550 nm thick epilayer at a substrate temperature of 700 °C. The thickness of the epilayer was confirmed by SIMS (Figure S7). The thicker layer enables capacitance-voltage (*C-V)* profiling, where the epilayer must be significantly thicker than the expected depletion width. Electrochemical capacitance-voltage (ECV) profiling was performed to measure the low unintentional carrier concentration as a function of depth. The carrier concentration as a function of depletion width was extracted assuming a dielectric constant of 11.5 for (100)-oriented $Ga_2O_3$ ($\epsilon_r$ = 10.2) miscut by 13.4° towards [-c].[23] The unintentional doping of the epilayer ranges from $2\times10^{15}$ $cm^{-3}$ – $7\times10^{15}$ $cm^{-3}$ as a function of depth, before steeply increasing towards the substrate which was Sn-doped with a free electron concentration of ~$5\times10^{18}$ $cm^{-3}$. The silicon trace measured by SIMS qualitatively follows the same trend as the unintentional doping profile. This is the lowest unintentional doping level reported for MBE-grown $Ga_2O_3$ to-date, the previous low concentration reported is $1\times10^{16}$ $cm^{-3}$ by the Air Force Research Laboratory.[74] We attribute the low background impurities to (1) use of ultra-pure 8N Ga source material, (2) operation of the plasma source at reduced power to prevent quartz bulb degradation,[74] (3) enhanced growth rates, and (4) a gate valve in front of the silicon cell to prevent any unintended evaporation. To confirm this result diodes were fabricated on an 11.1° substrate and a similarly low UID level of 3-$4\times10^{15}$ $cm^{-3}$ was observed (Figure S8).

Finally, we fabricated simple planar Schottky barrier diodes (SBD) with Pt Schottky contacts using MBE epilayers grown on 13.4° offcut substrates to measure leakage current and forward current density, noting that these films are highly insulating with a low unintentional doping level. The diode characteristics are shown in Figure 7a. The turn-on is sharp but not fully ideal, n = 1.37, and the leakage is flat to -20V where the electric field profile begins to punch into the highly doped substrate (the diode ideality fit is shown in Figure S9). The rectification ratio, taken at ±2V, is ~$10^5$. Capacitance-voltage measurements at 100 kHz were only able to deplete ~20 nm above the substrate but still show a carrier concentration of $3\times10^{18}$ $cm^{-3}$ in the substrate and $3\times10^{16}$ $cm^{-3}$ at a depth 20 nm above the substrate (Figure S10). The devices have no edge termination, passivation, dielectric, or electric field management applied. Despite the large leakage beyond -20V, some devices were driven to breakdown and demonstrate a breakdown up to 83 V (Figure 7b), corresponding to a critical field around 1.56 MV/cm. Previously reported (001) $Ga_2O_3$ SBDs with the same planar geometry and no edge termination show a critical field of ~ 1 MV/cm.[75]

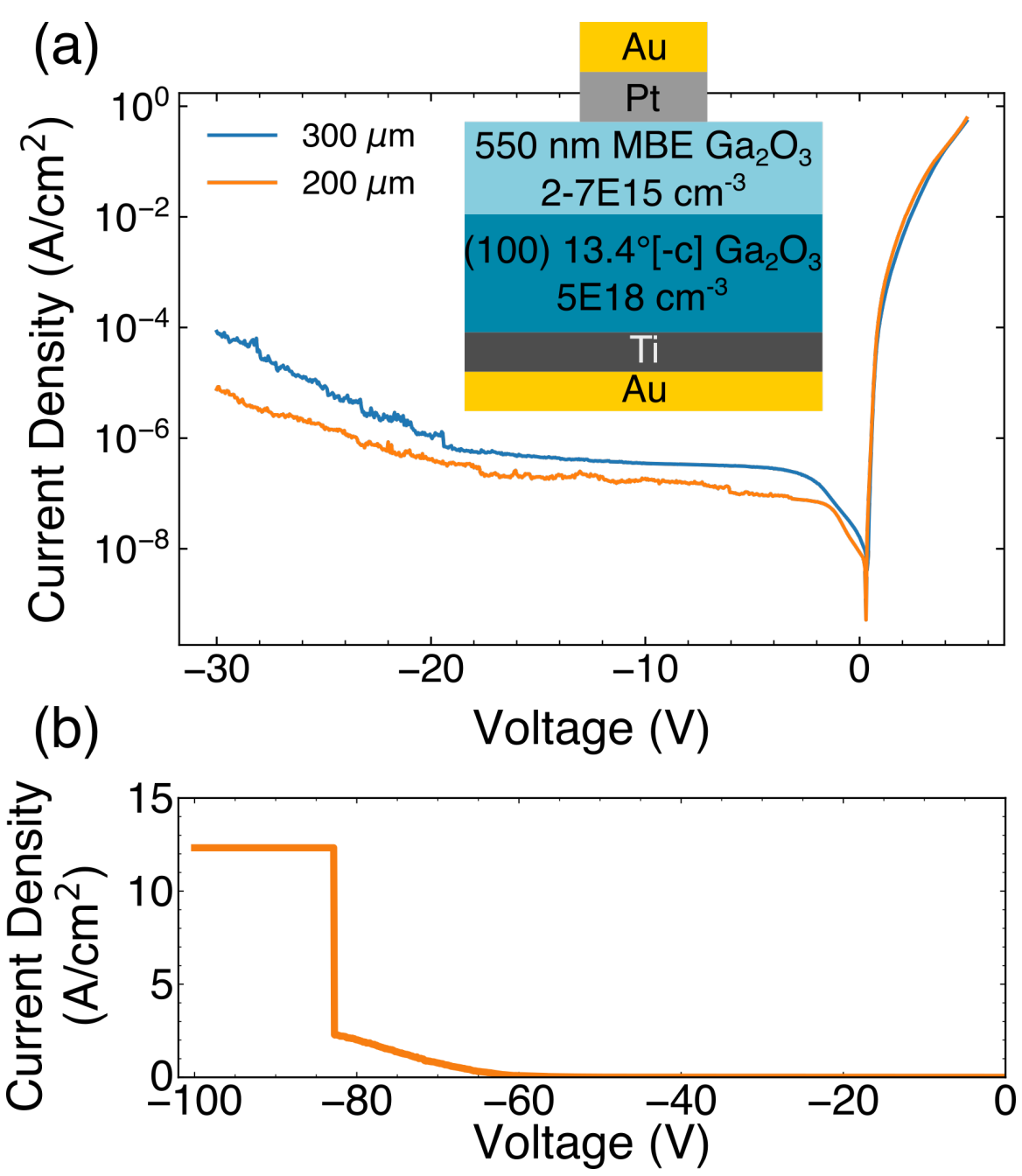

*Figure 7. **Current-voltage characteristics of highly offcut (100) $Ga_2O_3$ Schottky barrier diodes. (a)** Pt Schottky barrier diodes fabricated on 550 nm thick MBE-grown $Ga_2O_3$ epilayers using 13.4° offcut (100) substrates show strong rectifying behavior. These devices have no passivation, edge termination, or electric field management. **(b)** Reverse breakdown measurement of the 200 μm diode in (a) with an 83V breakdown corresponding to an average field ~1.56 MV/cm.*

## Conclusions

In conclusion, we have demonstrated that the growth rate of $Ga_2O_3$ thin films on (100) substrates can be completely recovered by employing large offcuts in the [00$\bar{1}$] direction. To avoid waste associated with grinding large offcuts, $Ga_2O_3$ ribbons were grown in by EFG by pulling in the [010] direction with an intentionally rotated seed. RHEED confirms the substrates have (100) terraces despite the large offcut angle. Growth rates up to 5.1 nm/min were demonstrated on 13.4° offcut wafers grown at 600 °C. The best surface morphology, associated with minimized step-bunching in the MBE-grown epilayers, was obtained at 650 °C substrate temperature for large offcuts (>10°) and 700 °C substrate temperature for moderate offcuts (~6°). High resolution electron microscopy demonstrated a continuous homoepitaxial interface with (100) terraces, corroborating step-flow growth. Electrical characterization of thick films shows a low unintentional doping density of $2\times10^{15}$ – $7\times10^{15}$ $cm^{-3}$ indicating the epilayers have low impurity concentrations, and these thick films maintain a low surface roughness of 1.3 nm RMS. Simple planar Schottky barrier diodes without edge termination or field management on highly offcut (100) $Ga_2O_3$ show similar turn-on and breakdown to diodes on other $Ga_2O_3$ orientations. Thus, highly offcut (100) $Ga_2O_3$ substrates are an effective platform to increase epitaxial growth rate and scale the wafer size for high-power and high-temperature electronic devices.

## Methods

### *Crystal growth*

The crystals were grown using an iridium crucible and die with an RF induction furnace operating at 10 kHz, in an inert atmosphere with a partial pressure of oxygen. The oxygen partial pressure was optimized to minimize the impact of decomposition of $Ga_2O_3$, while minimizing the oxidation of Ir components. The die was 2 mm thick by 57 mm wide, and ribbons were pulled up to 22 cm long. Raw materials used were high purity powders of gallium oxide ($Ga_2O_3$, 99.999%) and tin oxide ($SnO_2$, 99.99%), and ribbons were grown with bulk (average) Sn concentrations from 2-5 × $10^{18}$ $cm^{-3}$.

### *Substrate preparation*

After growth, the ribbons were planarized on both sides using a CNC mill. Then, each ribbon was mounted on a metal polishing plate and processed using the general procedure described by Lavelle et al.[76] First, a mechanical polishing step was implemented to remove a minimum of 60 μm from all locations using 70% concentrated Hyperion 1-3 μm diamond slurry on a Pureon SUBA™ 800 felt pad at a pressure of 2.0 psi. Then, two chemical-mechanical polishing (CMP) steps were utilized. After diamond mechanical polishing, 10 μm of material was removed using 50% concentrated Pureon ULTRA-SOL® 7H 0.07 μm colloidal silica slurry with a pH of 4 on a Pureon SUBA™ X felt pad. Finally, an additional 10 μm of material was removed using the same acidic colloidal silica slurry on a Pureon POLITEX™ poromeric pad. A pressure of 2.5 psi was used for both CMP steps and was decreased to 1.5 psi for the final 2 hours of polishing on the poromeric pad to mitigate subsurface damage. After CMP, the ribbons were mechanically diced into individual 10x10 mm parts using a DISCO DFD6240 dicing saw with a resin-bonded blade. An AFM image of a 5.7° [-c] offcut wafer polished in this manner is shown in Figure S2a.

*Epitaxial growth*

Epitaxial growth of $Ga_2O_3$ was performed in a Scienta-Omicron 3" oxide MBE on the previously described substrates. The wafers were solvent-cleaned using acetone, methanol, and isopropyl alcohol, and rinsed in deionized water. After the organic clean, the wafer was etched in 130 °C $H_3PO_4$ (85% bottle strength) for 15 minutes to remove ~100 nm from the surface, followed by a 15-minute room temperature surface clean using 2.5% TMAH (CD-26 developer) to remove remnants from the chemical mechanical polishing process (Figure S2b). The wafers were finally annealed at 950 °C for 3 hours in air to order the surface into steps (Figure S2c). The wafers were bonded to silicon handles using indium for introduction to the MBE chamber. The wafers were degassed at 750 °C for 10 minutes and annealed in an oxygen plasma before growth. To determine growth rate, an $(Al,Ga)_2O_3$ phasor layer was grown prior to the $Ga_2O_3$ layer for 90 s with an Al flux of 2 – 4E-8 torr beam equivalent pressure (BEP) to enable observation of X-ray diffraction (XRD) thickness fringes. Gallium and aluminum, of 8N and 7N purity respectively, were sourced from effusion cells. Oxygen was provided by a 13.56 MHz RF plasma source with a quartz bulb, and the operating conditions were set at 250 W and 3.0 SCCM. The growth chamber pressure during deposition was approximately 2.5E-5 torr. The silicon source is isolated behind a gate valve to prevent oxidation and unintentional doping, and the gate valve was closed during UID growth. The substrate temperature was monitored by a thermocouple, and we have found good temperature agreement with previous growth chambers calibrated by optical methods.[77] The growths were monitored by 20 kV reflection high-energy electron diffraction (RHEED) with differential pumping to protect the filament from oxidation.

*Epilayer Characterization*

High-resolution x-ray diffraction (HRXRD) analysis was performed using Cu Kα X-rays, collimated by a parabolic mirror, monochromated by a 4-bounce channel-cut Ge (200) crystal. The offcuts indicated were applied to the χ-axis, and the sample was rotated until the (400) diffraction peak was observed in a symmetric geometry. Atomic Force Microscopy (AFM) was performed using a Bruker Dimension Icon in tapping mode or ScanAsyst mode with an etched Al/Si tip with a ~2 nm radius. Scans were obtained across the step edges to improve resolution.

Specimens for scanning transmission electron microscopy (STEM) were prepared using standard FIB liftout procedures on a Thermo Fisher Scientific Helios 5 Hydra plasma-focused ion beam (PFIB) using Xe ions.[78] After PFIB thinning, the specimens were further thinned and polished using a Fischione 1040 NanoMill for the following steps: +/-10° tilt, 900 eV, 150 μA for 60 minutes on each side and +/-10° tilt, 500 eV, 120 μA for 30 minutes on each side. The specimens were imaged using a Thermo Fisher Scientific Spectra 200 STEM operating at 200 kV with a probe current of approximately 100 pA and convergence angle of 24.2 mrad providing a probe size of approximately 80 pm. The specimens were aligned to the $Ga_2O_3$ <010> zone axis for high-angle annular dark-field (HAADF) imaging and analysis.

Secondary ion mass spectrometry (SIMS) depth profile measurements were performed on a Cameca IMS-7f. Measurements were taken using both a cesium primary ion beam with negative secondary ions (15kV) and an oxygen primary ion beam with positive secondary ions (10kV).

*Electrical Measurements*

Electrochemical capacitance-voltage (ECV) profiling was performed with an electrolyte consisting of a 0.2M NaOH solution buffered with ethylenediaminetetraacetic acid (EDTA) at a frequency of 5.0 kHz. Electrical devices were fabricated after epitaxial growth. First, residual indium from bonding was removed with undiluted 36% HCl. Next, the planar back ohmic contact was deposited. First, the surface was exposed to 100 W UV Ozone for 10 minutes at room temperature. Next, 5 nm Ti and 100 nm Au were deposited by electron beam evaporation. Finally, the epitaxial wafer face was cleaned using the same UV Ozone recipe and a 20 nm Pt/100 nm Au Schottky contact was deposited through a shadow mask. The devices were probed with micromanipulators using a Keithley 4200 semiconductor parameter analyzer. The back-side of the wafer was contacted using conductive copper tape. *CV* data were obtained at 100 kHz with a delay factor of 1 s, a filter factor of 3 s, and an A/D aperture time of 1s.

**Acknowledgements**

This work was authored by the National Laboratory of the Rockies for the U.S. Department of Energy (DOE), operated under Contract No. DE-AC36-08GO28308. Funding is provided primarily by the Office of Critical Minerals and Energy Innovation (CMEI) Advanced Materials & Manufacturing Technologies suboffice (AMMTO), for the substrate growth and polishing, MBE growth, and microscopy measurements. Electrical characterization was supported as part of A Center for Power Electronics Materials and Manufacturing Exploration (APEX), an Energy Frontier Research Center funded by the U.S. Department of Energy,

Office of Science. The views expressed in the article do not necessarily represent the views of the DOE or the U.S. Government.

Supplementary Material for Fast Epitaxial Growth of $Ga_2O_3$ on Highly-Offcut (100)-Oriented Substrates

M. Brooks Tellekamp, Drew Haven, David Joyce, John Mangum, Kevin Schulte, Anna Sacchi, Matt Young, Andriy Zakutayev

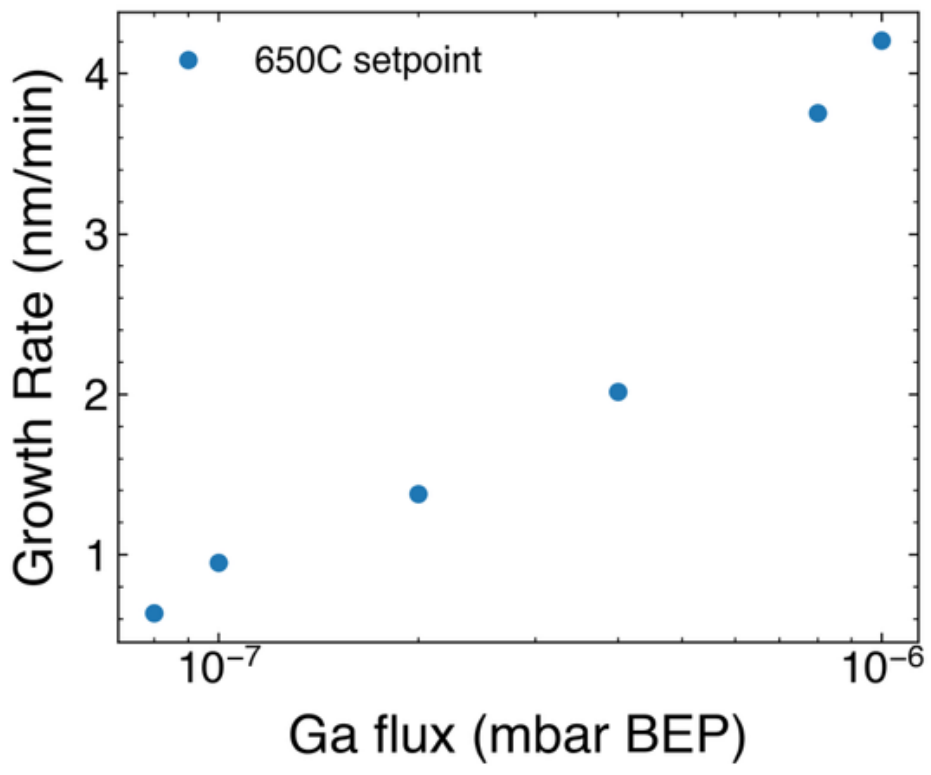

*Figure S1. Growth rate calibration at 650 °C, varying Ga flux, using (010)-oriented $Ga_2O_3$ substrates purchased from Novel Crystal Technology. (010) is known to have a faster growth rate and serves as a comparison point.*

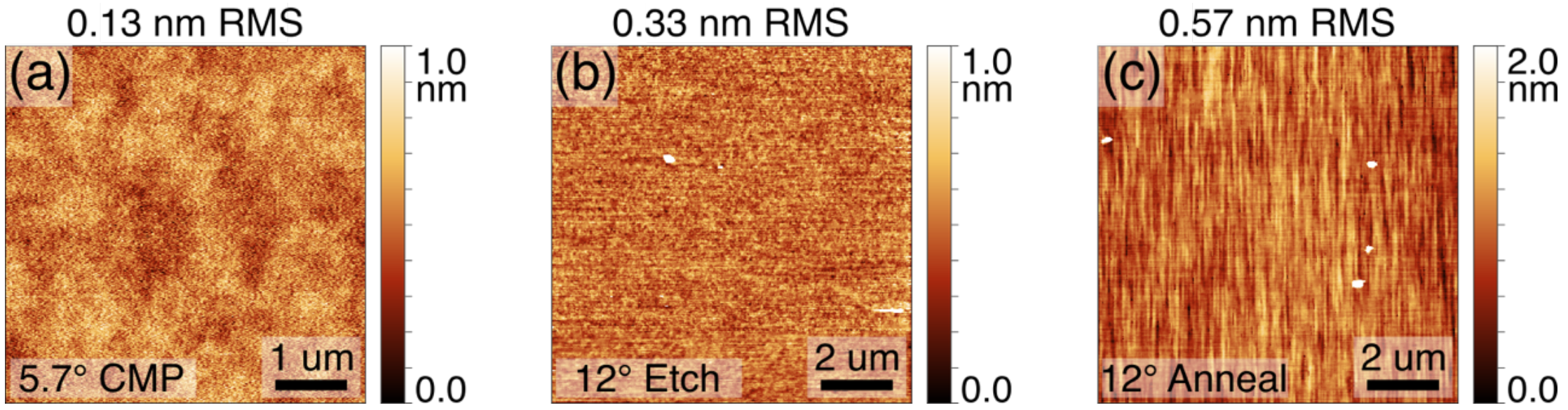

*Figure S2. Exemplary AFM of (a) the -5.7° offcut wafer after CMP, (b) a -12° offcut wafer after CMP and pre-growth wet etching, and (c) the same -12° offcut wafer after annealing at 950 °C in air to order the surface into steps and terraces.*

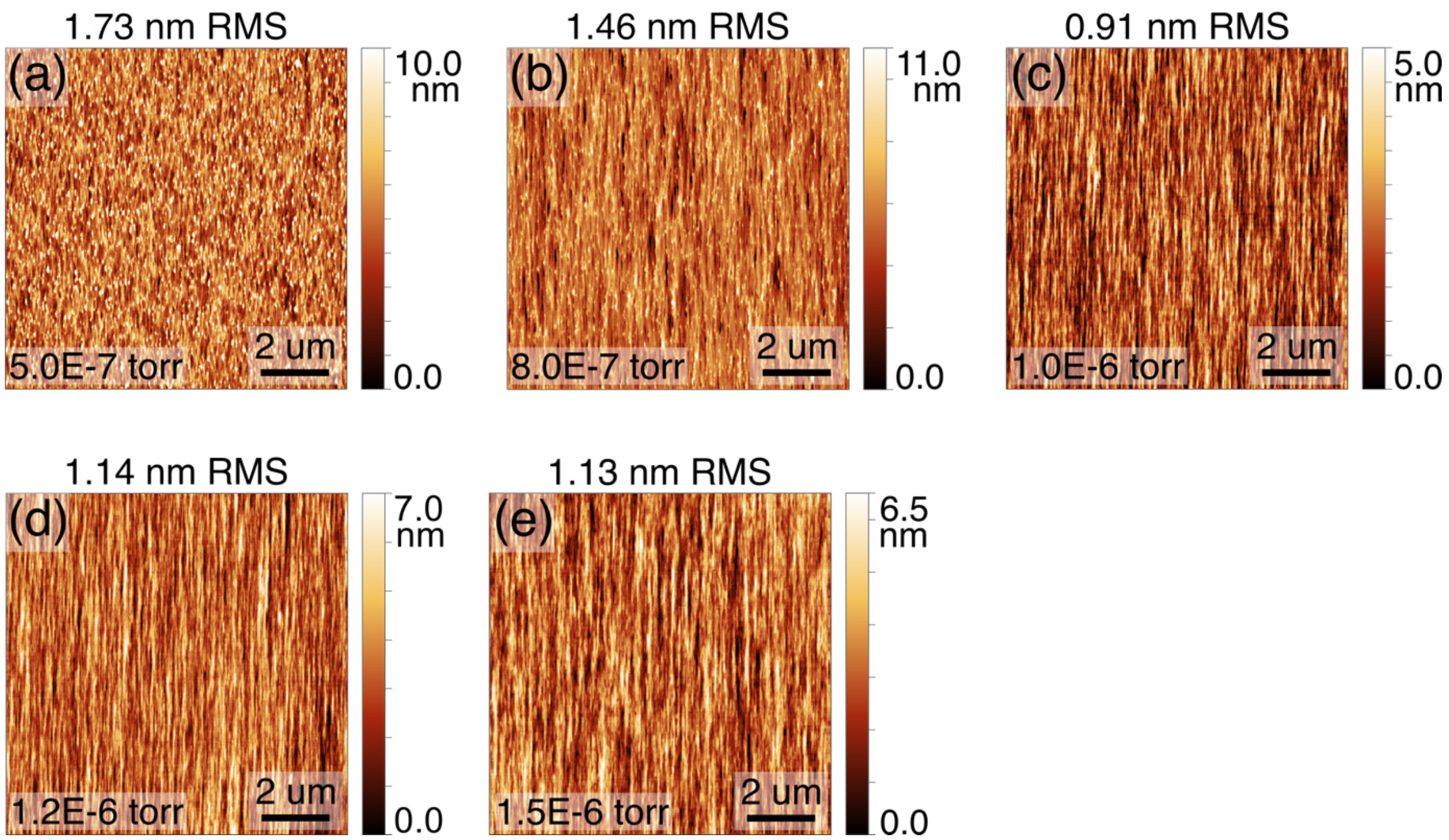

*Figure S3. AFM images of 100 nm $Ga_2O_3$ films from Figure 4a using a substrate temperature of 650 °C and 13.4° offcut substrates. Ga fluxes are given in the bottom left of the images. Lower fluxes promote the formation of triangular facets consistent in angle with (012) planes.*

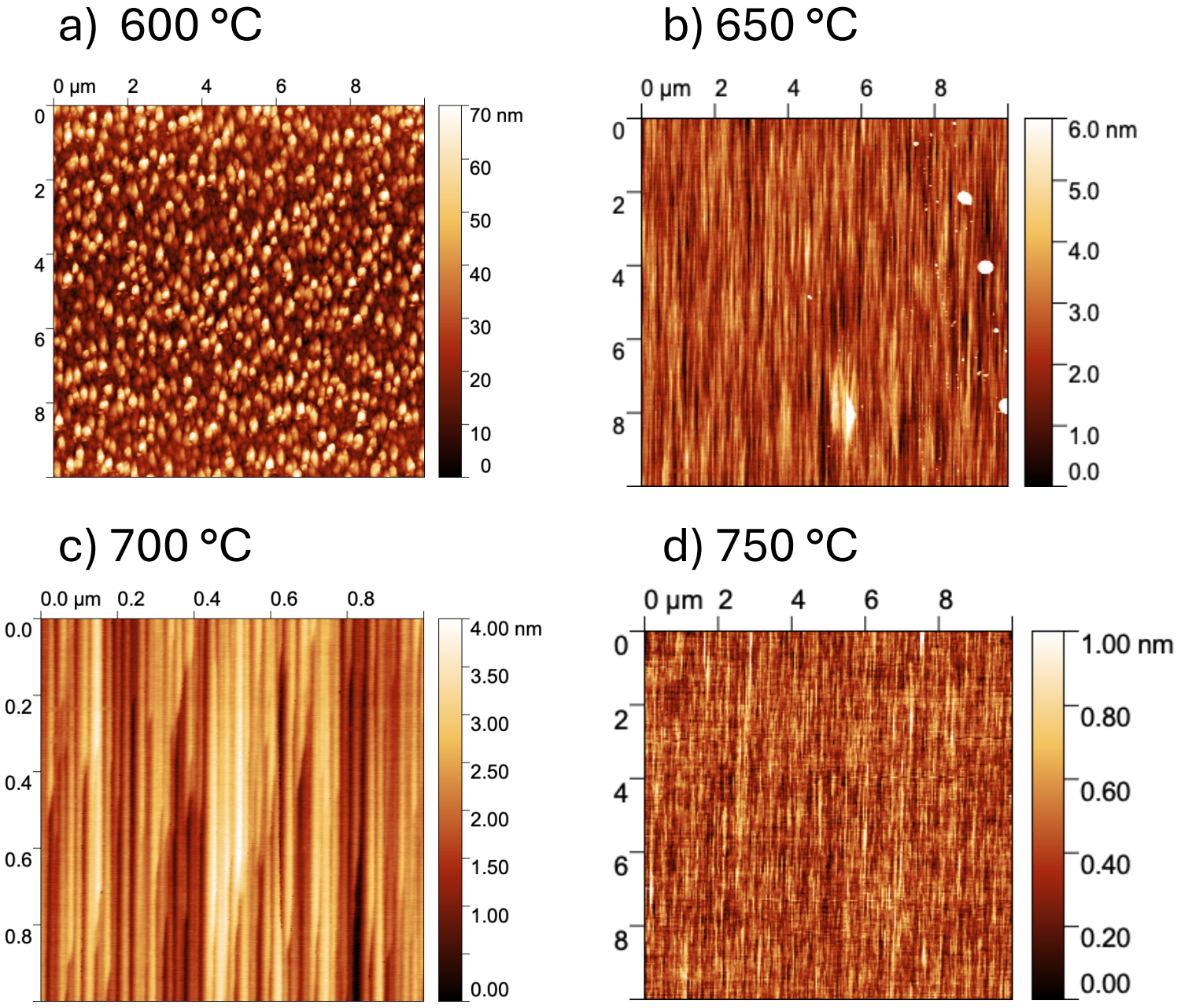

*Figure S4. AFM images of epilayers grown on 5.7° offcut wafers at substrate temperatures of (a) 600 °C, (b) 650 °C, (c) 700 °C, (d) 750 °C. The 750 °C sample is only ~10 nm thick, therefore was excluded from the surface roughness analysis in manuscript Figure 4a.*

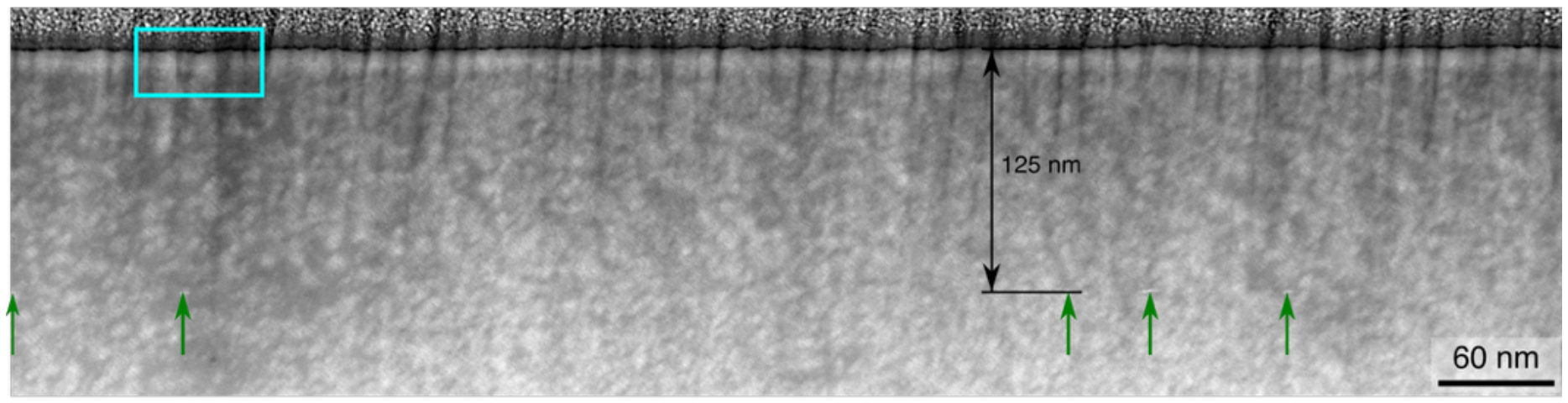

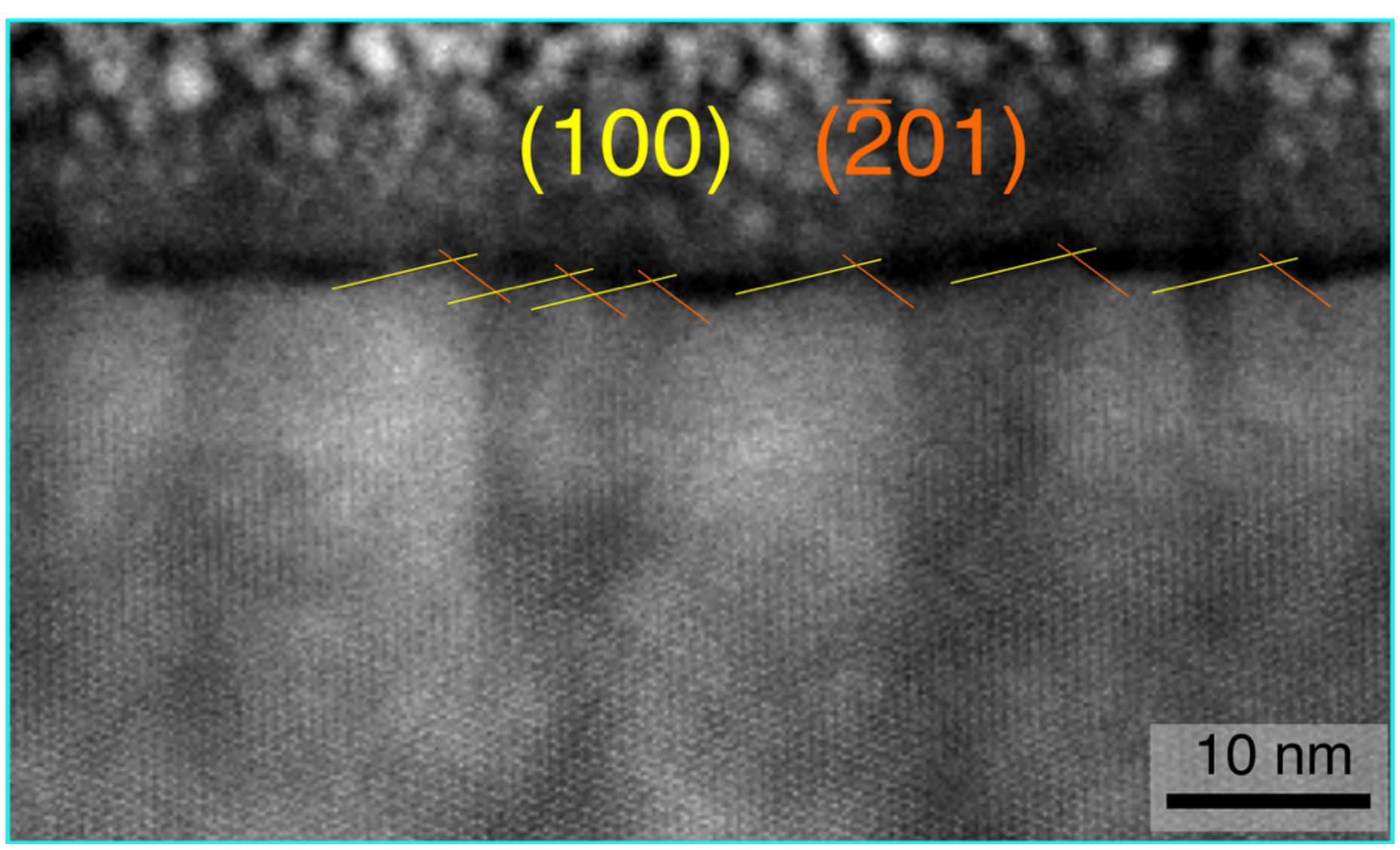

*Figure S5.* ***800 nm wide HAADF STEM image****. (a) Wide view STEM showing the epilayer thickness, where the interface was located from surface cleaning defects marked by green arrows. The location of (b) is marked in cyan. (b) zoomed in image of (a) marking (001) terraces in yellow and ($\bar{2}$01) step edges in orange.*

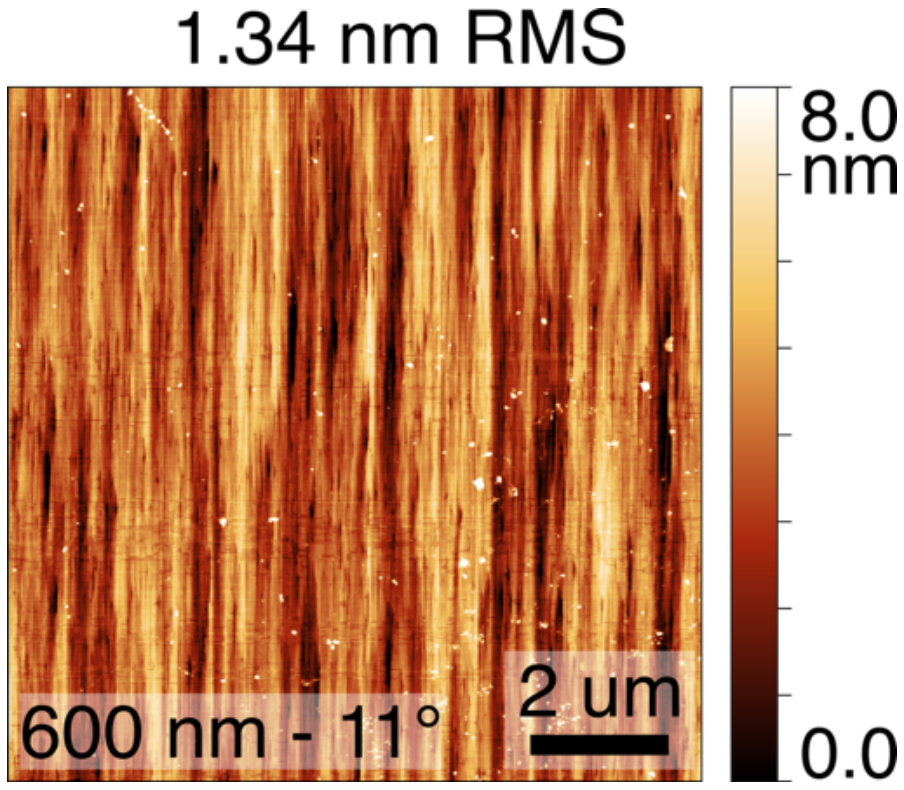

*Figure S6. AFM image of a 600 nm thick epilayer grown on an 11.1° offcut wafer shows maintained steps and a low 1.34 nm RSM surface roughness, indicating the step-flow growth mode is maintained over significant thickness.*

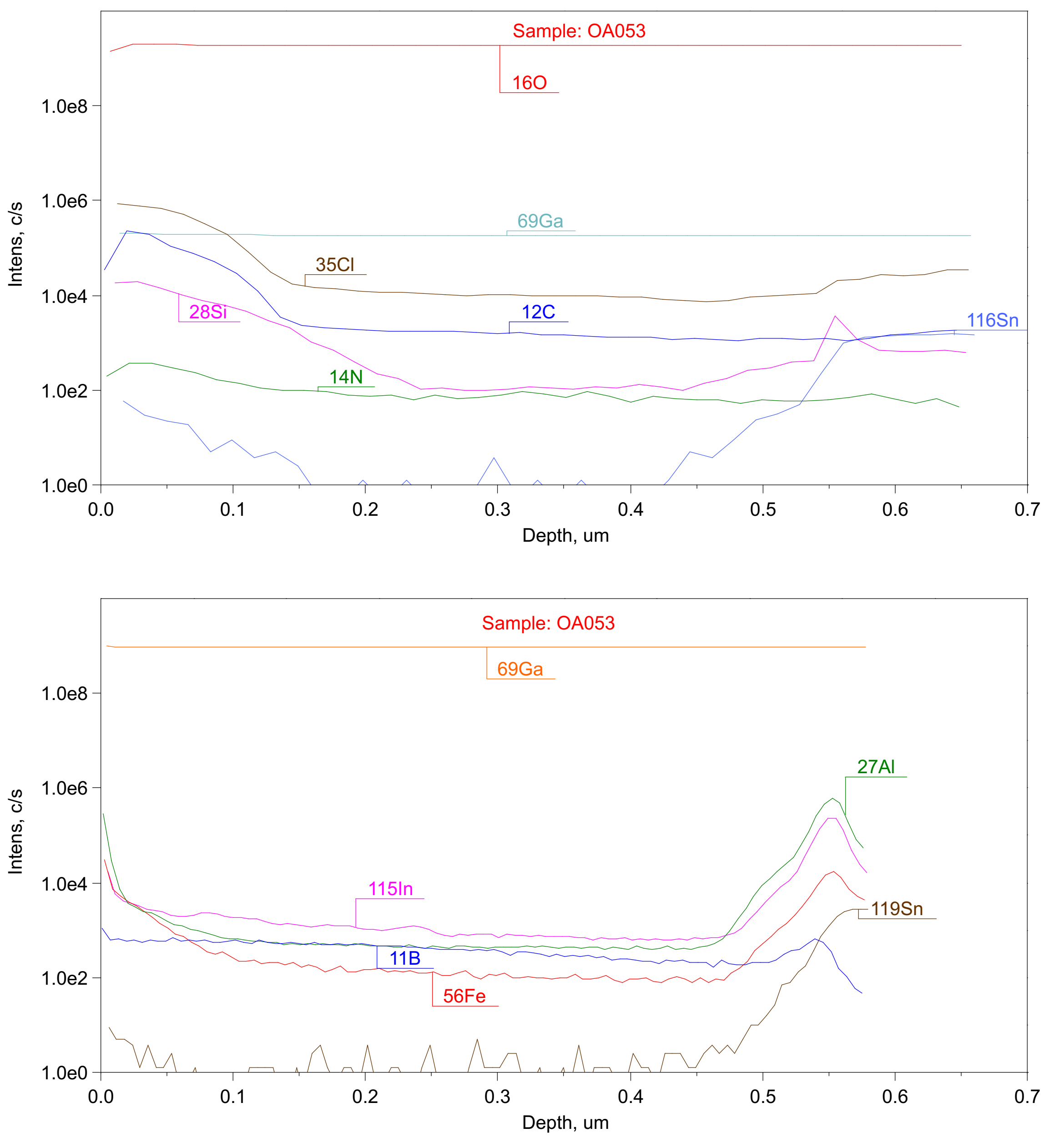

*Figure S7. Secondary ion mass spectrometry analysis of a 550 nm thick $Ga_2O_3$ epilayer grown at 700 °C on a 13.4° offcut (100) $Ga_2O_3$ substrate. (a) negative ions. (b) positive ions. The Si concentration mirrors the doping concentration measured by ECV in the manuscript Figure 6.*

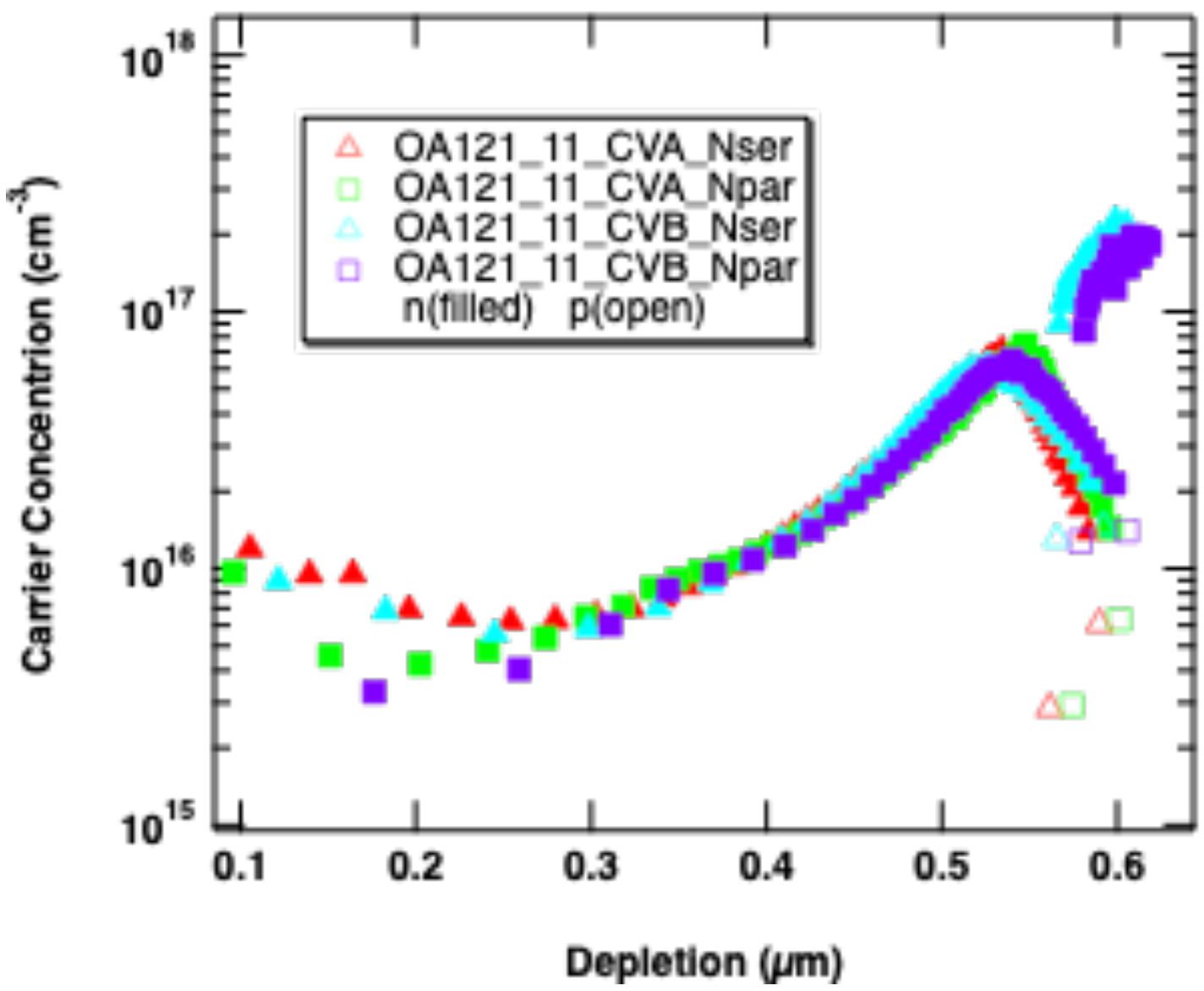

*Figure S8. Electrochemical capacitance-voltage on a 11.1° offcut MBE epilayer showing low unintentional n-type doping of 3-4E15 cm$^{-3}$ for the appropriate parallel model.*

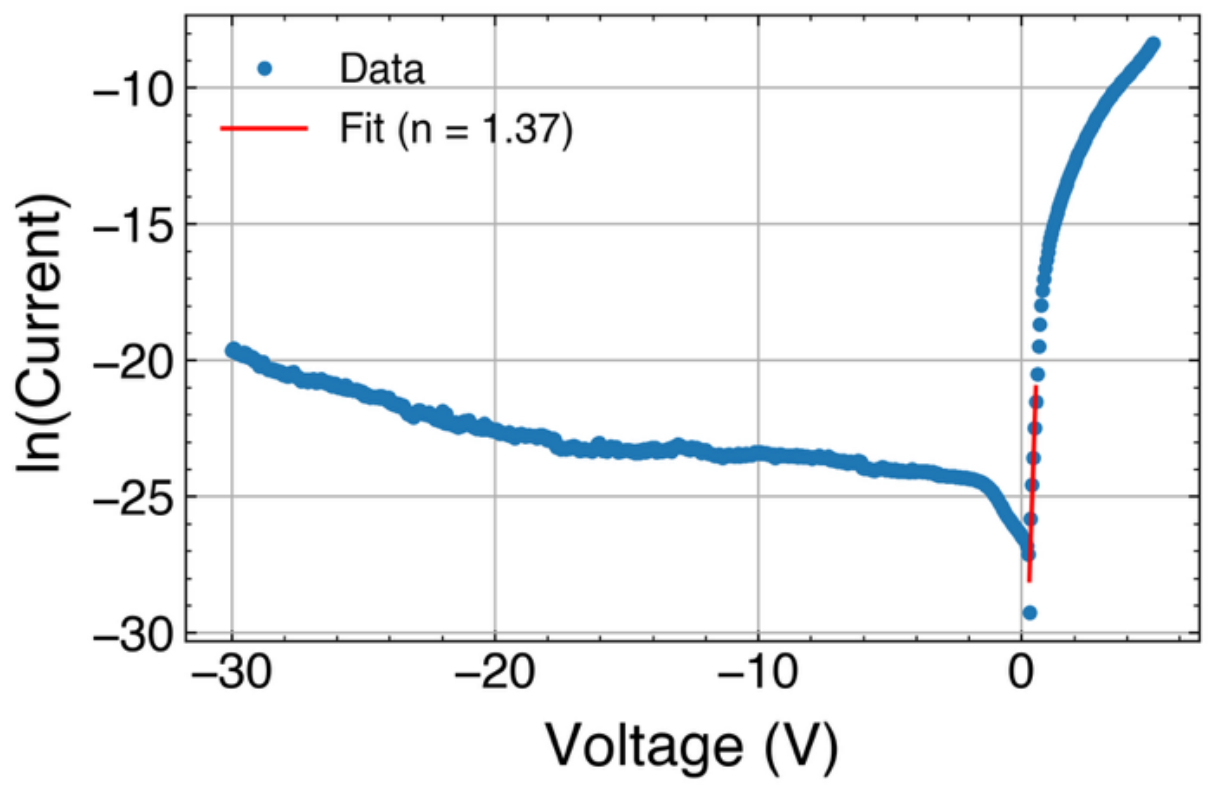

*Figure S9. Extracted ideality factor of 1.37 from a plot of ln(I) vs V.*

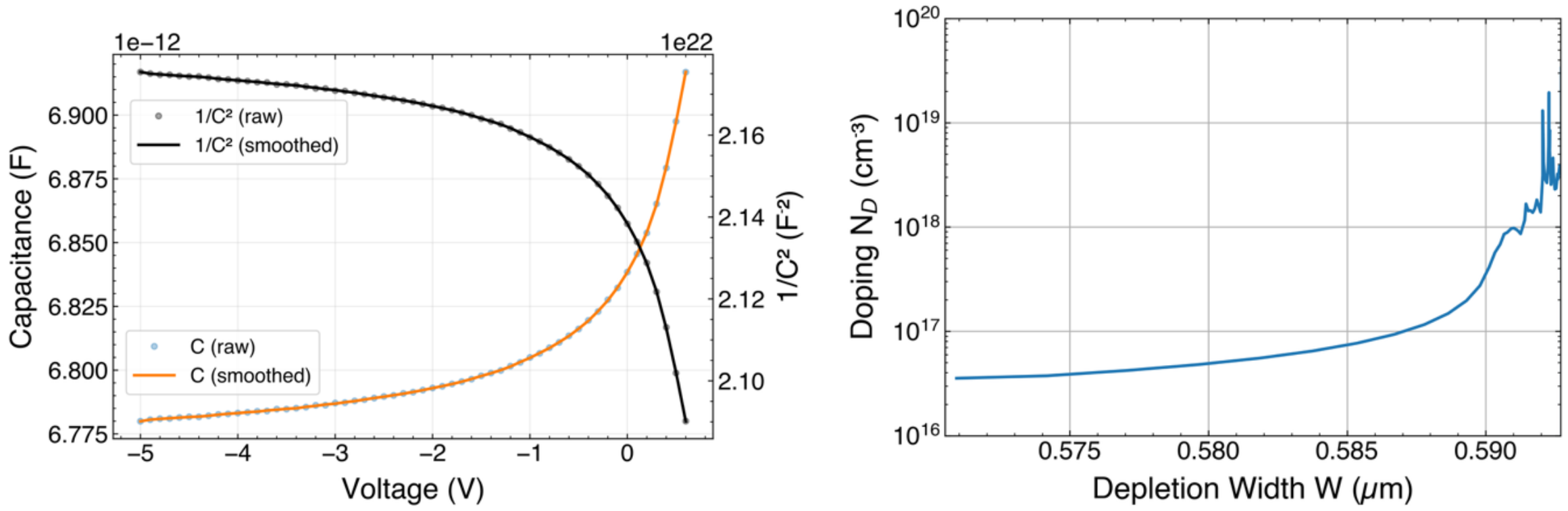

*Figure S10. Capacitance-voltage measurements of the SBD in manuscript Figure 7. (left) Capacitance versus voltage and $1/C^2$ vs voltage measured at 100 kHz with a small signal amplitude of 100 mV. (b) Carrier concentration vs depletion width. The depletion measured before diode turn-on is only within 20 nm of the substrate and not representative of the bulk. Despite this, the unintentional doping level is still < 5E16 $cm^{-3}$ at 10 nm above the substrate and the doping in the substrate is correctly measured at ~3E18 $cm^{-3}$.*